\documentclass[
 reprint,
 amsmath,amssymb,
 aps,
prb,
superscriptaddress,
showpacs, showkeys]{revtex4-1}

\usepackage{color}
\usepackage{graphicx}
\usepackage{dcolumn}
\usepackage{bm}
\usepackage{hyperref}
\hypersetup{bookmarksnumbered, pdfpagemode=UseOutlines, 
colorlinks=true, citecolor=blue, filecolor=blue, linkcolor=blue, urlcolor=blue}

\graphicspath{X:\paper
}

\begin{document}


\title{Evolution of the magnetoresistance lineshape with temperature and electric field across  Nb-doped SrTiO$_3$ interface }

\author{A. Das}
\thanks{corresponding author}
\email{arijit.das@rug.nl}
\author{S. T. Jousma}
\author{A. Majumdar}
\author{T. Banerjee}
\thanks{corresponding author}
\email{t.banerjee@rug.nl}
\affiliation{University of Groningen, Zernike Institute for Advanced Materials, 9747 AG Groningen, The Netherlands}

%


\date{\today}

\begin{abstract}
We report on the temperature and electric field driven evolution of the magnetoresistance lineshape at an interface between Ni/AlO$_x$ and Nb-doped SrTiO$_3$. This is manifested as a superposition of the Lorentzian lineshape due to spin accumulation and a parabolic background related to tunneling anisotropic magnetoresistance (TAMR). The characteristic Lorentzian line shape of the spin voltage is retrieved only at low temperatures and large positive applied bias. This is caused by the reduction of electric field at large positive applied bias which results in a simultaneous reduction of the background TAMR and a sharp enhancement in spin injection. Such mechanisms to tune magnetoresistance are uncommon in conventional semiconductors.
	
\end{abstract}

\keywords{Complex Oxides, Spin injection, electric field, Tunneling Anisotropic Magnetoresistance (TAMR)}
\maketitle

Spin voltage measured at different semiconducting interfaces have been widely studied using different combination of materials as spin contacts and employing different measurement techniques.  Such studies are commonly performed using the popular three terminal (3T) and four terminal non-local (NL) geometries\cite{Lou2007,Jonker2007,Dash2009,Reyren2012,Kamerbeek2015,Lee2017}. In spite of the fact that both these electrical transport schemes fail to resolve outstanding issues related to the precise understanding, origin and magnitude of spin accumulation across semiconducting interfaces, these are more accentuated using the 3T geometry\cite{Deranlot2009,Txoperena2014,Park2015,Pu2016,Inoue2015,Park2015}. This is rooted in the inability of the 3T scheme to clearly ascertain the origin and magnitude of the spin voltage, possible considerations being spin accumulation in the semiconductor or localized states either in the tunneling barrier or at the semiconducting surface \cite{Song2014}. \\
Earlier studies involving amorphous tunnel barriers showed that the tunneling conductance and spin polarization can be strongly influenced by the presence, concentration and type of impurities in the tunneling barrier via the formation of impurity mini bands and highly conducting multiresonant channels\cite{Tsymbal2009,Jansen2000,Jansen2004}. Attempts to mitigate such impurities by designing epitaxial barriers has also proved non-trivial in this context\cite{Reyren2012,Inoue2015}. Additionally, the nature and type of impurities offer further challenges to validate proposed theories that seek to explain experimental observations using either spin injection (ferromagnet/tunnel barrier) or non-magnetic (metal/tunnel barrier) contacts\cite{Txoperena2014,Park2015}.  Increasing the parameter space by using new transport schemes and/or choosing different materials will be an useful approach to understand the experimental findings related to the origin of spin accumulation across semiconducting interfaces. 
One such material interface that enables tunability of electronic properties, relevant for spin transport is that of complex oxides \cite{Sulpizio2014}. Although such material interfaces are commonly replete with oxygen vacancies and surface charge\cite{Janotti2014,Saraf2016,Andra2017}, the tunability of several functional properties with temperature, electrical field, stress and strain has led to the unexpected emergence of new phenomena not encountered in other material systems. \\
In this context SrTiO$_3$ (STO) is a relevant material. STO single crystals exhibit a large dielectric permittivity ($\epsilon_r$) at room temperature, that is anisotropic in different crystalline directions and increases non-linearly with temperature, electric field and frequency\cite{Neville1972,Yamamoto1998,Spinelli2010}. Doping of Nb, La at the Ti site transforms it into a degenerate ionic semiconductor (n-doped) with unconventional charge transport characteristics, triggered by the strong temperature and electric field dependence of the intrinsic dielectric permittivity\cite{Rana2012,Spinelli2010}.Additionally, the broken inversion symmetry at the surface of STO leads to Rashba spin-orbit fields\cite{Nakamura2012,Khalsa2013} which when tuned by electric fields either at the interface of a 2-DEG (LAO-STO) or at the interface of Nb-doped SrTiO$_3$ (Nb:STO) with Co/AlO$_x$ results in tuning of spin transport parameters as demonstrated in recent works\cite{Reyren2012,Kamerbeek2015}. \\
\begin{figure*}[]
	\centering
	\includegraphics[scale=0.7]{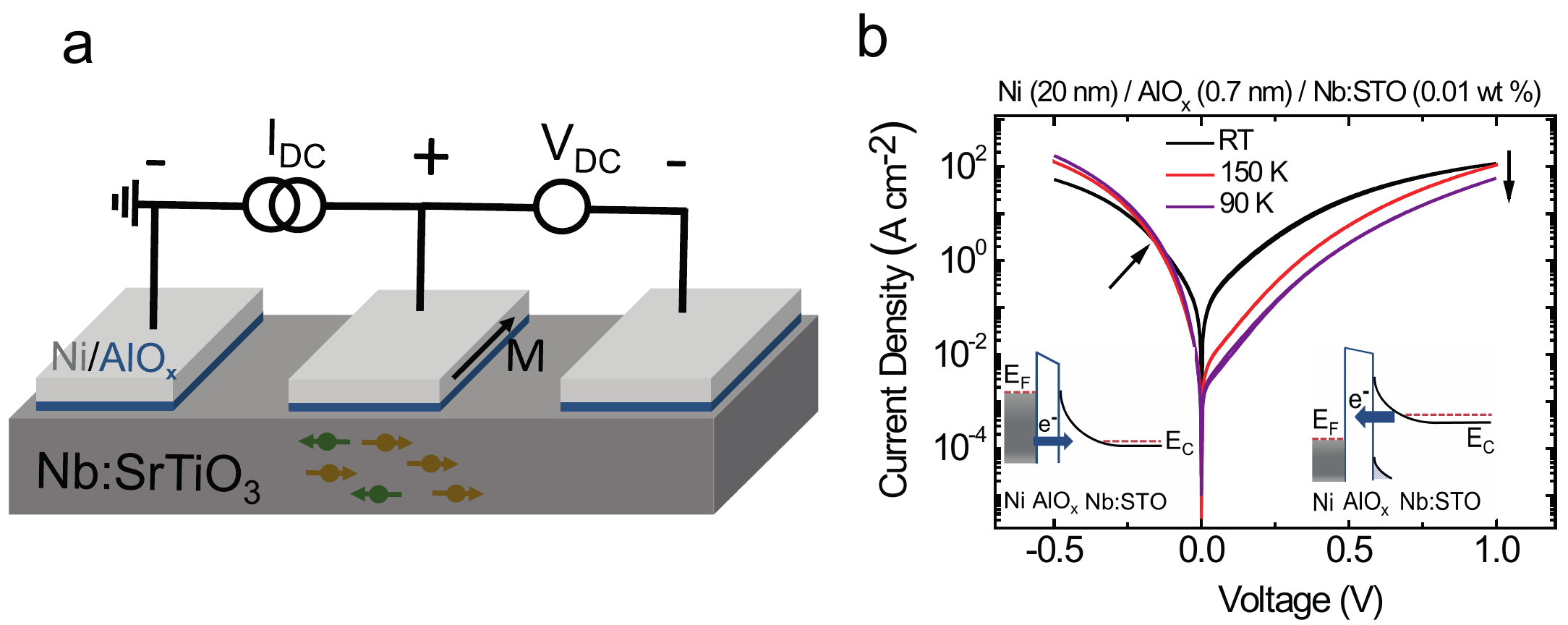}
	\caption{
(a) Electrical measurement scheme using a three terminal (3T) geometry with Ni (20 nm)/AlO$_x$ (7\AA) spin injection contacts on Nb:STO (0.01 wt\% Nb). A constant current (I$_{DC}$) is sourced across the central spin contact producing a non-equilibrium spin accumulation in Nb:STO that is probed by a voltage (V$_{DC}$) across the same contact. (b) Charge transport characteristics (J-V) plotted at three different temperatures ( RT, 150 K and 90 K). In forward bias, the arrow that points downward indicates decreasing forward current on reducing temperature. In reverse bias, the arrow indicates the point at which the reverse current at low temperatures crosses the current at room temperature indicating increasing current on reducing temperature. The inset to this figure shows the change in the potential landscape across the Schottky interface of Nb:STO with applied bias (right for forward bias, left for reverse bias).} 
\end{figure*} 
We report on an unconventional magnetoresistance response at the spin injection interface of Ni/AlO$_x$ on Nb:STO at room temperature, as observed in the lineshape of the spin voltage. The lineshape is found to be a superposition of different magnetoresistance effects across the Schottky interface between Nb:STO and the spin contact, not reported earlier across such or other semiconducting interfaces. The superimposed signals are predominantly related to spin accumulation either at the semiconductor or due to the localized states and a background Tunneling anisotropic magnetoresistance (TAMR). TAMR arises due to spin orbit coupling (SOC) effects in Nb:STO and is exhibited as a change in the tunneling conductance (spin voltage) when rotating the magnetization of the ferromagnet with respect to the current flow direction\cite{Kamerbeek2018}. A systematic study of the temperature and bias dependence of the electronic transport across the Nb:STO interface reflects that the non-linear response of the intrinsic dielectric permittivity in Nb:STO strongly influences the lineshape of the measured spin voltage. We find that enhanced tunneling at low temperatures and the concomitant reduction of the electric field at large positive bias leads to the recovery of the conventional Lorentzian line shape of the spin voltage with a simultaneous reduction of the background TAMR. \\
In this work, we design a spin injection interface by evaporating spin contacts of Ni/AlO$_x$ on a low doped Nb:STO semiconductor with a Nb doping of 0.01 wt\% (N$_d$=3x10$^{18}$ cm$^{-3}$) \cite{Rana2012} and perform electrical three terminal studies on the fabricated devices. We have used single crystalline 0.01 wt\% Nb:STO obtained from Crystec GmbH and use standard chemical protocol to prepare single terminated surface consisting of TiO$_2$ planes\cite{Koster2012}. A 7 $\AA$  thin film of Al is deposited using electron beam evaporation on the surface of STO followed by an in-situ plasma oxidation. Finally, 20 nm of Ni is evaporated followed by 20 nm of Au as a capping layer forming an interface of Au/Ni/AlO$_x$ on Nb:STO. The sample was then patterned using UV lithography and dry etching into contact pillars of junction area ranging from 50 $\mu$m to 200 $\mu$m x 200 $\mu$m. Figure 1a shows the 3T device geometry where current I$_{DC}$ is sourced across the central spin contacts that consists of Ni/AlO$_x$ on Nb:STO. The voltage drop across the interface is measured as V$_{DC}$ across the same central contact. The charge transport (J-V) characteristics for such spin contacts is shown in Fig. 1b, where the current density is plotted with respect to the applied voltage bias at three temperatures (RT, 150 K and 90 K). The J-V characteristics are dominated by transport across the Schottky interface as is clear from the variation of the current in the forward bias with decreasing temperature. Transport is governed by thermally assisted field emission into Ni/AlO$_x$ contact across the Schottky barrier (forward bias) in Nb:STO. The forward bias transport is not significantly influenced by the electric field reduction of the increased $\epsilon_r$ in Nb:STO at lower temperatures. On the other hand, an increase in the reverse current with reduction in temperature indicates larger tunneling to occur at reverse bias. This is expected at a Schottky interface with Nb:STO since a strong increase in $\epsilon_r$ with decreasing temperature gives rise to a steeper band bending at reverse bias due to an increase in the built-in electric field\cite{Kamerbeek2014,Kamerbeek2015a}. Thus, in these fabricated devices, the potential landscape is engineered such that the charge (and spin) transport is governed by the tunneling conductance both across the thin tunneling barrier of AlO$_x$ as well as the Schottky interface at Nb:STO.\\
\begin{figure}[]
	\centering
	\includegraphics[scale=0.35]{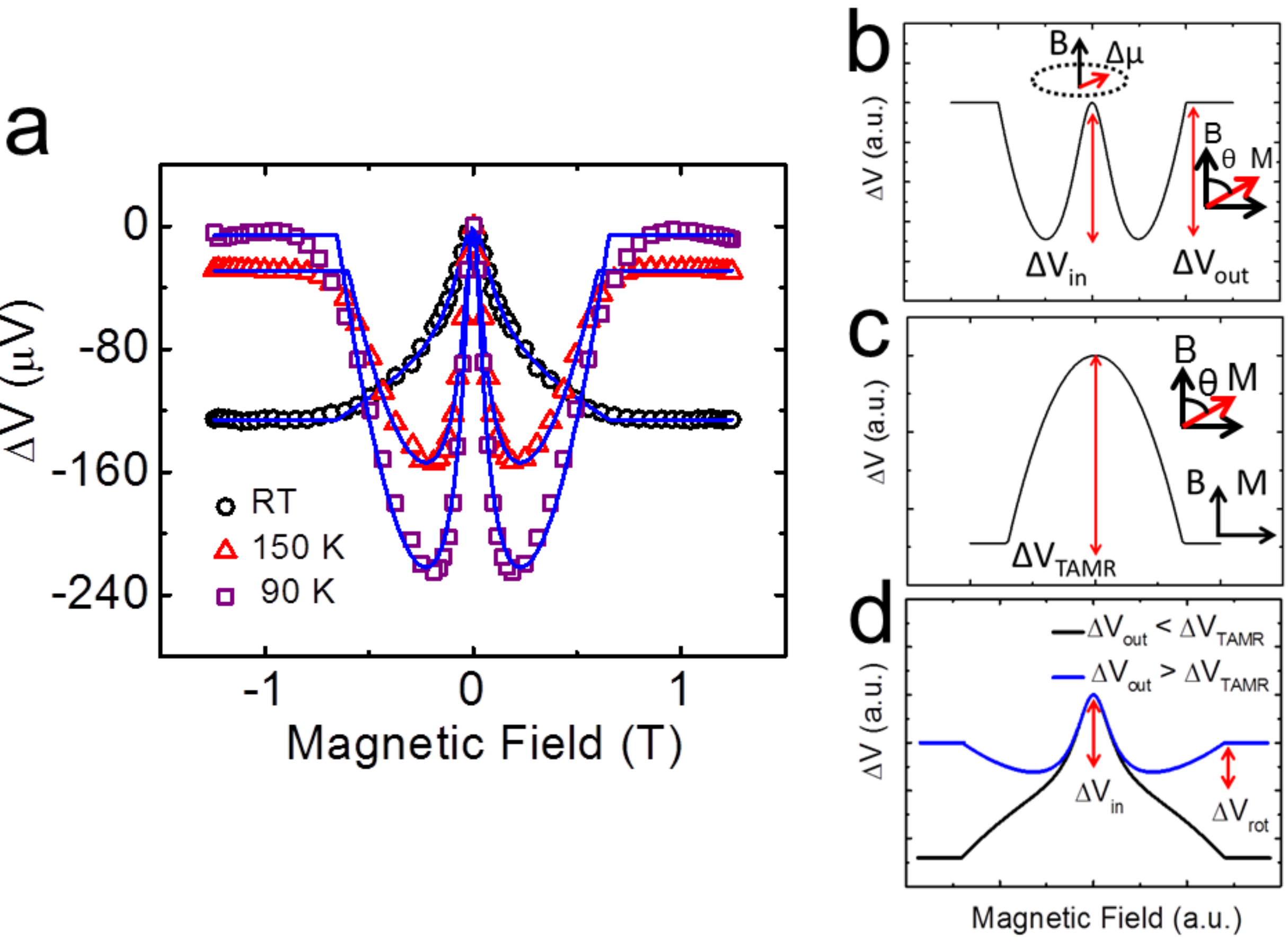}
	\caption{
(a) Spin voltage responses are shown at three different temperatures (RT, 150 K and 90 K) with an out-of-plane magnetic field at a fixed junction voltage V$_b$ = +1 V. The solid blue lines are fits using Eq. 1. (b) Spin voltage is simulated with out-of-plane magnetic field using Eq. 1. The conventional Lorentzian response is indicated by $\mathrm{\Delta V_{in}}$ and the out-of-plane spin accumulation by $\mathrm{\Delta V_{out}}$. (c) Parabolic TAMR response with an out-of-plane magnetic field is simulated. The difference in the tunnelling resistance when the magnetization of the ferromagnet rotates from in-plane to out-of-plane is given by $\mathrm{\Delta V_{TAMR}}$. (d) Competition between TAMR and spin accumulation results in complete suppresion of $\mathrm{\Delta V_{out}}$ when TAMR response is larger as shown by black line whereas larger spin accumulation response partially suppress $\mathrm{\Delta V_{out}}$ as shown by blue line. 
	} 
\end{figure} 
Using such engineered interfaces, we measure spin voltages with a magnetic field applied perpendicular to the device interface. Figure 2a shows the response at three different temperatures (RT, 150 K and 90 K). As stated earlier, a constant current bias I$_{DC}$ is sourced across the central spin contact (Fig. 1a) and a voltage V$_{DC}$ is measured across the same contact. By subtracting the charge related background corresponding to the junction voltage V$_b$, the spin voltage $\Delta$V is obtained. The spin voltage signals ($\Delta$V) are obtained at a fixed junction voltage (V$_b$) of +1 V. We observe clear contrast in the lineshape of the magnetoresistance signals with temperature as shown in Fig. 2a. The signal decreases with an unconventional Lorentzian lineshape with increasing field strength at room temperature and saturates at field values between 650-700 mT corresponding to the saturation magnetization in Ni. The spin voltages at lower temperatures also saturate around the same point but shows a strong upturn before saturation. The dephasing of the spins occur due to an increase in the spin precession amplitude with an increasing out-of-plane magnetic field at a Larmor frequency given by  $\omega_{L}$. This gives rise to a Lorentzian lineshape of the spin voltage and the Bloch equation that describes the effect of spin dephasing and 1-D diffusion across the contact is given by:
\begin{equation} 
\Delta{V} = \Delta{V_{rot}}\cos^{2}{\theta} + \frac{\Delta{V_{in}}\sin^{2}{\theta}}{\sqrt{2}}\sqrt{\frac{1+\sqrt{1+(\omega_{L}\tau_{in})^2}}{1+(\omega_{L}\tau_{in})^2}}
\label{Eq:1}
\end{equation} \\
$\mathrm{\Delta V_{in}}$ is the spin voltage due to the spin dephasing effect describing the amplitude of the Lorentzian spin signal. $\mathrm{\tau_{in}}$ is the spin lifetime that depends on the inverse of the full width half maximum (FWHM) of the Lorentzian signal, $\mathrm{\theta}$ is the angle between the magnetization of Ni with surface normal and $\mathrm{\Delta V_{rot}}$ is the spin voltage that arises due to the rotation of the magnetization of the ferromagnet. The solid blue lines in Fig. 2a are fits to the spin voltage response  at three temperatures using Eq.1.\\
Figure 2b shows the simulated spin voltage responses, $\Delta$V, as described by the model above. It represents the Hanle effect with the conventional Lorentzian decay and an upturn due to the gradual rotation of the magnetization in Ni out-of-the-plane followed by a saturation when both the spin accumulation and the magnetization are perpendicular to the applied field. The amplitude of the signals for in and out-of-plane spins are given by $\mathrm{\Delta V_{in}}$ and $\mathrm{\Delta V_{out}}$ respectively. Such a response is reflected in the measured $\Delta$V at 90 K (Fig. 2a) where the amplitudes of $\mathrm{\Delta V_{in}}$ and $\mathrm{\Delta V_{out}}$ are similar. A partial suppression of the signal upturn due to $\mathrm{\Delta V_{out}}$ is observed at 150 K (Fig. 2b) indicating the presence of a second spin response. The latter is also responsible for a complete suppression of $\mathrm{\Delta V_{out}}$ at RT and results in an additional linewidth broadening and an unconventional Lorentzian response at RT. Figure 2c shows the simulated response of tunneling anisotropic magnetoresistance (TAMR) and is parabolic with the out-of-plane magnetic field. As mentioned earlier, the tunneling conductance changes when the magnetization in Ni rotates from in-plane to out-of-plane with respect to the applied current direction resulting in such a lineshape. The saturation response has the same origin as discussed for Fig. 2a. When the two effects are in competition, the resultant shape of the spin voltage can be very different as shown in Fig. 2d. The lineshape in black corresponds to the case when the TAMR response is dominant over the spin accumulation, resulting in complete suppression of $\mathrm{\Delta V_{out}}$. The lineshape in blue represents the case when spin accumulation response is dominant over TAMR (the signal due to $\mathrm{\Delta V_{out}}$ increases). The interplay between $\mathrm{\Delta V_{out}}$ and $\mathrm{\Delta V_{TAMR}}$ is represented by $\mathrm{\Delta V_{rot}}$ (the first term in Eq.1) and is responsible for the rotation of the magnetization in Ni. Thus the observed change in the lineshape of the  magnetoresistance with temperature as shown in  Fig. 2a is due to the competition of the out-of-plane spin accumulation and the TAMR response. On reducing the temperature, an enhancement in the spin accumulation signal is observed that overshadows the dominance of the TAMR effect at room temperature (Temperature dependence of the spin voltage response at different bias can be found in the accompanying supplementary information Fig. S2). Such an unconventional lineshape of the spin voltage at RT that evolves to a Lorentzian, at lower temperatures, has not been observed in other semiconducting interfaces and underpins the role of the engineered interface.  \\
\begin{figure}[]
\centering
\includegraphics[scale=0.9]{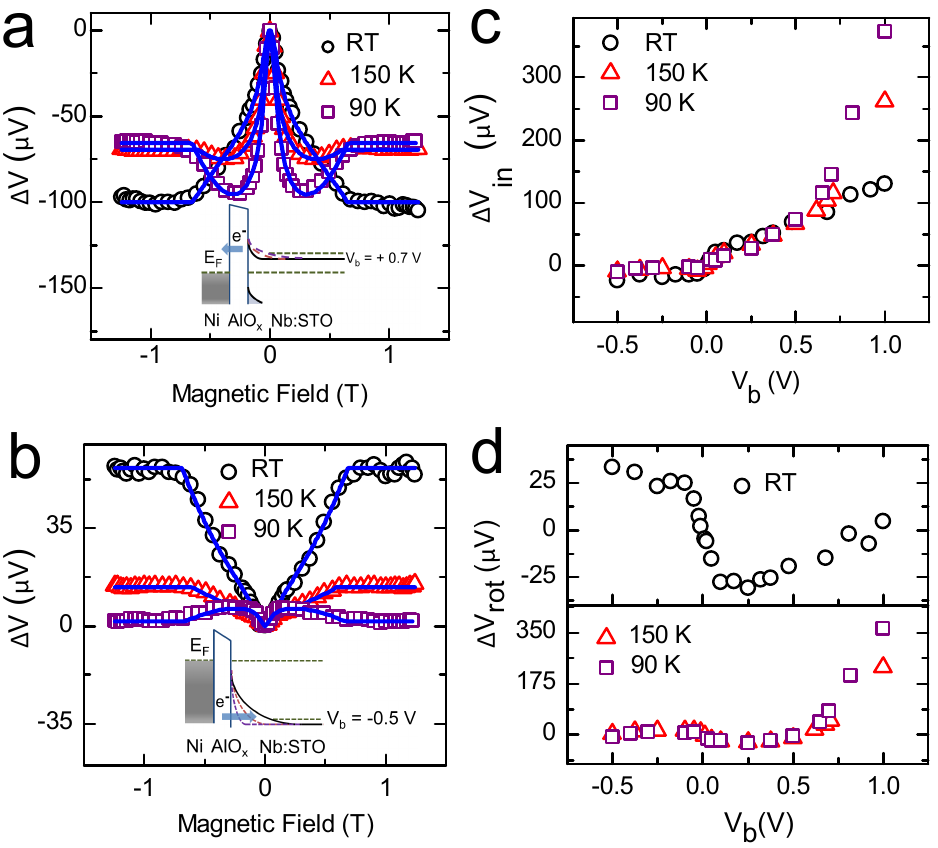}
\caption{
(a) Spin voltage responses at three different temperatures (RT, 150 K and 90 K) with an out-of-plane magnetic field at a fixed junction voltage V$_b$ = +0.7 V (forward bias). The inset in the bottom is a schematic of the potential profile for increasing forward bias. The black solid line is for room temperature and the black dashed line is for low temperatures (b) Spin voltage response for reverse bias at a fixed junction voltage V$_b$ = -0.5 V. The inset in the bottom represents the potential profile for increasing reverse bias- all lines and colors have the same meaning as that of the inset in (a). The solid blue lines are fits using Eq. 1. in both cases. (c) Bias dependent variation of $\mathrm{\Delta V_{in}}$ at three temperatures (RT, 150 K and 90 K). (b) Bias dependent variation of $\mathrm{\Delta V_{rot}}$ at room temperature (top panel) and for low temperatures (90 K and 150 K) (bottom panel)}
\end{figure}
To understand the influence of the engineered Nb:STO interface on the spin transport further, we analyse the temperature and bias dependence of the lineshapes of the spin voltages. Shown in Fig. 3a and b are for forward bias (V$_b$ = +0.7 V) and reverse bias (V$_b$ = -0.5 V) of the Schotkky interface. We extract $\mathrm{\Delta V_{in}}$ and $\mathrm{\Delta V_{rot}}$ and plot their variation with applied bias for the three different temperatures. 
The spin dephasing parameter represented by $\mathrm{\Delta V_{in}}$ is plotted with junction voltage V$_b$, obtained from the current bias I$_{DC}$ as shown in Fig. 3c. The interplay of the TAMR and out-of-plane spin accumulation response, given by $\mathrm{\Delta V_{rot}}$ and plotted at three different temperatures is shown in Fig. 3d. In Fig. 3c we observe an increase of $\mathrm{\Delta V_{in}}$ at all temperatures by increasing the junction voltage from negative to positive. For V$_b$, between -0.5 V to +0.5 V (regime i), the spin voltage $\mathrm{\Delta V_{in}}$ with bias is similar for all the three temperatures, whereas for V$_b$ beyond 0.5 V (regime ii), $\mathrm{\Delta V_{in}}$ signal sharply increases with reducing temperature. Similarly, $\mathrm{\Delta V_{rot}}$ also shows a strong variation with bias at the three temperatures as shown in Fig. 3d. At RT (top panel in Fig. 3d), the variation in $\mathrm{\Delta V_{rot}}$ is distinctly different from that at lower temperatures (regime i) as shown in the bottom panel. In regime (i), the strong presence of TAMR, due to large electric fields at reverse bias, overshadows the contribution of $\mathrm{\Delta V_{out}}$, leading to an increase in $\mathrm{\Delta V_{rot}}$ with increasing reverse bias at room temperature.  At low temperatures,  $\mathrm{\Delta V_{rot}}$  approaches to zero (since the TAMR effect decreases) at high reverse bias. At low forward bias (regime i), $\mathrm{\Delta V_{rot}}$ is negative (larger TAMR response) and gradually approaches to zero (regime (ii)) at room temperature, where the spin accumulation, $\mathrm{\Delta V_{in}}$, increases to larger positive values at low temperature. Although we cannot independently disentangle the contribution due to out-of-plane spin accumulation and TAMR response, the bias modulation of  $\mathrm{\Delta V_{rot}}$  and $\mathrm{\Delta V_{in}}$ allows us to understand their interdependence, albeit qualitatively, from their variation with temperature.\\
The lineshape of the Hanle curves and the bias variation of the different components in $\Delta$V, at different temperatures can be reconciled if we look at the factors that govern the potential landscape of the engineered interface. The insets in Fig. 3a and b shows the schematic of the potential landscape dominated by the Schottky interface at Nb:STO for increasing forward and reverse bias. At room temperature, the band gradually flattens at higher forward bias (solid black curve) reducing the contribution due to the Schottky interface. The bias dependence of the Schottky profile modulates the built-in electric field at this interface and this tunes the Rashba spin-orbit field, that arises at the surface of STO due to broken inversion symmetry\cite{Kamerbeek2015}. For increasing applied forward bias, the electric field decreases, thus the contribution due to $\mathrm{\Delta V_{in}}$ is markedly prominent over that of $\mathrm{\Delta V_{rot}}$ indicating a reduced TAMR effect. This increment, at larger bias (regime ii) becomes quite clear at lower temperatures where the contribution of TAMR is negligible, due to decreasing  electric fields. This enhances the contribution of spin conserving tunneling processes, as found by an increase in $\mathrm{\Delta V_{in}}$ in Fig. 3c at larger applied bias. On the other hand, with increasing reverse bias Rashba SOC increases due to an increase in the electric field. This results in a larger TAMR response as evident in Fig. 3d, masking any spin voltage signals related to spin accumulation $\mathrm{\Delta V_{in}}$ in (Fig. 3c). Thus the different role of the electric field driven effects on $\mathrm{\Delta V_{in}}$ and $\mathrm{\Delta V_{rot}}$ at different temperatures controls the evolution of the lineshape in the spin voltage response across such interfaces, and is unlike that reported across interfaces with conventional semiconductors such as Si \cite{Dankert2013}. \\ 
Engineering the potential landscape at the Nb:STO interface and thus the interplay between the electric field and temperature dependence of the Rashba SOC results in the evolution of the Lorentzian lineshape of the observed spin voltage. At room temperature the built-in electric field at the dominant Schottky interface enables the observation of a large TAMR effect, whereas reducing the temperature changes the transport regime to that of a dominant tunneling one enabling the observation of a spin voltage related to spin accumulation. Such effects are strongly manifested in Nb:STO due to the additional tunability of the dielectric permittivity in Nb:STO and cannot be observed in conventional semiconductors. This additional flexibility in device design at such material interfaces leads to new understanding on the origin of the different contributions to the spin voltages measured using the 3T electrical transport scheme. \\

See Supplementary Material for the temperature dependence of the spin voltage response at different bias.\\

The authors would like to thank A.M. Kamerbeek for his helpful insight and discussion.  AD and TB also thanks J.G. Holstein and H.M. de Roosz for the technical support. This work is supported by the Dieptestrategie grant 2014 from Zernike Institute for Advanced Materials, University of Groningen.
                             
%

\end{document}